\documentclass[aps,prb,twocolumn,superscriptaddress]{revtex4}

\usepackage[dvips]{graphicx}
\usepackage{color}

\newcommand{\HTO}{\text{Ho$_2$Ti$_2$O$_7$} }

\begin{document}

\title{Dynamics of the Magnetic Susceptibility Deep in the Coulomb Phase of the Dipolar Spin Ice Material Ho$_2$Ti$_2$O$_7$}

\author{J.~A.~Quilliam} \altaffiliation{Present address: Laboratoire de Physique des Solides, Universit\'{e} Paris-Sud 11, UMR CNRS 8502, 91405 Orsay, France}
\author{L.~R.~Yaraskavitch}

\affiliation{Department of Physics and Astronomy and Guelph-Waterloo Physics Institute, University of Waterloo, Waterloo, ON N2L 3G1 Canada}
\affiliation{Institute for Quantum Computing, University of Waterloo, Waterloo, ON N2L 3G1 Canada}

\author{H.~A.~Dabkowska}
\affiliation{Brockhouse Institute for Materials Research, McMaster University, Hamilton, ON, L8S 4M1, Canada}

\author{B.~D.~Gaulin}
\affiliation{Brockhouse Institute for Materials Research, McMaster University, Hamilton, ON, L8S 4M1, Canada}
\affiliation{Department of Physics and Astromony, McMaster University, Hamilton, ON, L8S 4M1, Canada}
\affiliation{Canadian Institute for Advanced Research, 180 Dundas Street West, Toronto, Ontario, Canada M5G 1Z8}

\author{J.~B.~Kycia}
\affiliation{Department of Physics and Astronomy and Guelph-Waterloo Physics Institute, University of
Waterloo, Waterloo, ON N2L 3G1 Canada}
\affiliation{Institute for Quantum Computing, University of
Waterloo, Waterloo, ON N2L 3G1 Canada}

\date{\today}

\begin{abstract}

Low-temperature measurements of the ac magnetic susceptibility along the $[110]$ direction of a single crystal of the dipolar spin ice material Ho$_2$Ti$_2$O$_7$, in zero static field, are presented.  While behavior that is qualitatively consistent with previous work on Ho$_2$Ti$_2$O$_7$ and the related material Dy$_2$Ti$_2$O$_7$ is observed, this work extends measurements to appreciably lower temperatures and frequencies.  In the freezing regime, below 1 K, the dynamics are found to be temperature activated, thus well described by an Arrhenius law with an activation energy close to $6J_\mathrm{eff}$, a result that is not easily explained with the current model of magnetic monopole excitations in dipolar spin ice.  The form and temperature dependence of the ac susceptibility spectra are found to be nontrivial and distinct from standard glassy relaxation.  Particular attention has been paid to correcting for the demagnetization effect, which is quite significant in these materials and has important, even qualitative, effects on the susceptibility spectra.

\end{abstract}


\maketitle


\section{Introduction}

Recent years have seen a flurry of activity stemming from the recognition~\cite{Castelnovo2008} that the unusual ground-state correlations of dipolar spin ice should support excitations that behave as magnetic charges.  These magnetic charges, or deconfined monopoles, are formed through the fractionalization of a single spin-flip (dipole) excitation.  By virtue of the underlying correlations of the dipolar spin ice `vacuum', the monopoles experience a Coulomb force between them and are found to flow under the influence of the magnetic field.~\cite{Bramwell2009}  Several experimental works and theoretical analyses of previous experimental work~\cite{Bramwell2009, Morris2009, Fennell2009,Kadowaki2009,Jaubert2009,Jaubert2010} have since verified the existence of these monopole excitations.  Furthermore, lithographically patterned, artificial spin ice systems have been created\cite{Wang2006} and also shown to exhibit monopole defects.\cite{Ladak2010, Mengotti2010, Morgan2010}  In particular, the slow relaxation that is observed in the ac magnetic susceptibility of spin ice~\cite{Snyder2001, Snyder2003, Snyder2004, Matsuhira2001, Matsuhira2000, Shi2007, Ehlers2004} has recently been attributed to the freezing out of dynamics of deconfined magnetic monopoles.~\cite{Jaubert2009}

Here, we present detailed low temperature ac susceptibility measurements, $\chi(f)$, on single crystal Ho$_2$Ti$_2$O$_7$. These measurements represent an important step in the study of monopole physics of dipolar spin ice for several reasons.  (i) Ho$_2$Ti$_2$O$_7$ has received less attention, at least from thermodynamic measurements, than Dy$_2$Ti$_2$O$_7$.  \HTO ac susceptibility measurements thus far have been limited to temperature scans of $\chi$ in higher temperature and frequency regimes.~\cite{Matsuhira2000, Cornelius2001, Ehlers2004} (ii) Our experiments represent the first low temperature ac susceptibility measurements performed on single crystal spin ice in zero static field along the [110] direction. (iii) In previous work~\cite{Matsuhira2001,Snyder2001, Snyder2003, Snyder2004, Shi2007} on Dy$_2$Ti$_2$O$_7$, there has not been a clear accounting of the demagnetization correction, which turns out to be rather important for quantitatively determining the intrinsic time constant of the system as a function of temperature and matching to theoretical predictions.\cite{Jaubert2009}  The demagnetization correction also has qualitative effects on the shape of the frequency spectra. (iv) This work extends measurements to much lower frequencies than previous work on any variety of spin ice, therefore delving deep into the Coulomb phase.  It is our hope that it will provide an important benchmark by which to test theories of monopole physics in dipolar spin ice.


In the pyrochlore materials Ho$_2$Ti$_2$O$_7$ (HTO) and Dy$_2$Ti$_2$O$_7$ (DTO), the magnetic ions Ho$^{3+}$ and Dy$^{3+}$ occupy a lattice of corner sharing tetrahedra.  The crystal field acting on those ions creates a strong Ising anisotropy, with the easy axis, known as the local [111] axis, pointing directly in or out of the tetrahedra.  For ferromagnetic nearest neighbor interactions, the spins are highly frustrated and possess a macroscopically degenerate set of ground states, consisting of two spins pointing into and two spins pointing out of each tetrahedron.~\cite{Harris1997}  This situation is directly analogous to the proton bonds in water ice, hence the name `spin ice'.  In both spin and water ice, the macroscopic degeneracy of ground states gives rise to an extensive residual entropy $S_0 = {R/2}\ln(3/2)$,\cite{Ramirez1999,denHertog2000,Higashinaka2003,Hiroi2003} known as Pauling's entropy.\cite{Pauling}  In HTO and DTO, the nearest neighbor (n.n.) ferromagnetic interaction is a result of dipole-dipole interactions, thus these systems are often referred to as dipolar spin ice.~\cite{denHertog2000}  Considering the antiferromagnetic n.n. exchange interaction $J$ and the Ising anisotropy of the spins, the relevant effective n.n. interaction becomes $J_\mathrm{eff} S_i^z S_j^z$ where
\begin{equation} J_\mathrm{eff} = 	m^2(5D - J)/3, \end{equation}
$m$ is the magnetic moment of the Dy$^{3+}$ or Ho$^{3+}$ ions and $D$ is the strength of the n.n. dipolar interaction.  The further neighbor dipolar interactions are very well screened in this system.~\cite{Gingras2001,Castelnovo2008}  Nonetheless, they are found, theoretically, to select a unique ground state for the system and should, in principle, result in a sharp, first-order phase transition.~\cite{Melko2001}  However, no such transition has been found in experiment,~\cite{Gardner2010,Fukazawa2002,Melko2004} possibly because of the difficulties of reaching equilibrium at temperatures well below 0.5 K.

The unusual disordered ice-like state of dipolar spin ice, with its two-in, two-out tetrahedra, can be thought of as a divergence-free $\nabla\cdot\vec{H} = 0$ vacuum, with fluctuating magnetic field lines.~\cite{Balents2010}  The simplest excitation out of this vacuum is the flipping of one spin.  This single spin flip affects two neighboring tetrahedra resulting in one with 3-in, 1-out and the other with 1-in, 3-out.  This removes the divergence-free character of the system and leads to two magnetic charges, or monopoles, of opposite sign, centered on those tetrahedra.  In the simple nearest neighbor model of spin ice, these monopole excitations are, once created, free to roam about on the lattice without any energy cost.  In dipolar spin ice, however, the longer range interactions give rise to a Coulomb interaction, $\sim 1/r$, between magnetic monopoles.\cite{Castelnovo2008,Jaubert2009,Jaubert2010}

This has been associated recently~\cite{Jaubert2009, Jaubert2010} with the slow relaxation observed in the ac magnetic susceptibility of spin ice.~\cite{Snyder2001, Snyder2003, Snyder2004, Matsuhira2001, Matsuhira2000, Shi2007, Ehlers2004}  At temperatures ranging from about 2 K to 6 K, where the monopole density is high but there are very few double defects (4-out or 4-in), DTO ac susceptibility data from Ref.~\onlinecite{Snyder2004} is well parametrized by an Arrhenius law with
	\begin{equation} \tau = \tau_0 \exp( E_A/T) = \tau_0 \exp( 2J_\mathrm{eff}/T),
	\label{ArrheniusLaw}
	\end{equation}
where $2J_\mathrm{eff}$ is the energy cost of a single monopole defect.  As the temperature is lowered and the concentration of monopoles decreases, the relaxation becomes slower than predicted by Eq.~\ref{ArrheniusLaw}.  A reasonably good fit of experimental data is obtained by performing Monte Carlo simulations of
a gas of monopole charges on the diamond lattice in the grand canonical ensemble with a largely temperature independent chemical potential $\mu$.~\cite{Jaubert2009,Jaubert2010}

Despite this strong verification of the monopole hypothesis, there remains some discrepancy between theory and experiment at lower temperatures (below $\sim 1$ K) in that the experiments find slower relaxation than is predicted by theory.\cite{Jaubert2010}  It is, thus, important to further explore the low temperature freezing of spin ice systems.  Ensuring an accurate correction of the demagnetization effect and taking measurements to lower frequencies, as performed here, will permit careful comparison of experimental results with the theory of monopole defects.  In contrast to the predictions of theory, our results show a temperature activated regime where the dynamics are described well by a surprisingly simple Arrhenius law with $6J_\mathrm{eff}$ activation energy at low temperatures.  This study, performed on Ho$_2$Ti$_2$O$_7$, as opposed to the more commonly studied Dy$_2$Ti$_2$O$_7$, also permits the observation of monopole physics with a different set of Hamiltonian parameters, namely the spin flip rate, effective n.n. interaction, $J_\mathrm{eff}$, and the monopole charge, $Q$.

\begin{figure}
\begin{center}
\includegraphics[width=3.25in,keepaspectratio=true]{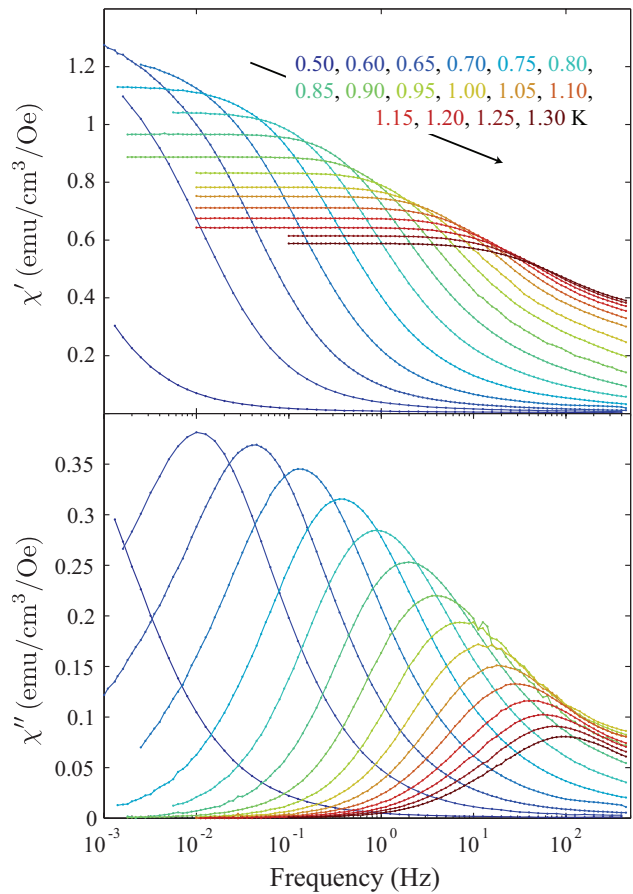}
\caption{(color online) Frequency scans of $\chi'$ and $\chi''$ of \HTO with the magnetic field aligned along the $[110]$ crystal orientation.  The data shown here has been corrected for demagnetization and was taken on a needle shaped sample (sample 1).  Slow relaxation is observed, characterized by a suppression of $\chi'$ at higher frequencies and a broad peak in $\chi''$.}
\label{ChiNeedleFigure}
\end{center}
\end{figure}

\section{Experiment}

The single crystal \HTO samples studied here were prepared at McMaster University.  The samples are from the same crystal growth employed in the neutron scattering studies reported in Ref.~\onlinecite{Clancy2009}.  They are grown by a floating-zone image furnace technique, the details of which are provided in Ref.~\onlinecite{Gardner1998}.  Two sample geometries were measured, with dimensions $1.1\times 1.1\times 2.6$ mm$^3$ and $0.6\times0.6\times 3$ mm$^3$ respectively.  In each case, the longest side of the crystal was the [110] crystal direction, also the direction of the applied ac magnetic field.


Susceptibility, $\chi$, measurements were performed using a magnetometer based on a superconducting quantum interference device (SQUID), mounted on a dilution refrigerator.  The second-order gradiometer sensing coils are contained in the bore of an excitation coil consisting of 375 turns of NbTi wire wound on a phenolic form with which an applied ac magnetic field, not exceeding 20 mOe, is generated.  The gradiometer is coupled to the SQUID by means of a superconducting flux transformer.  Further refinement of the magnetometer balance is achieved with a trim coil coupled to one branch of the gradiometer, in parallel with the excitation coil.  The SQUID is contained within a small lead shield, and the entire gradiometer within another, larger lead shield.  Further shielding is provided by a cryogenic $\mu$-metal shield surrounding the vacuum can of the cryostat and by a room-temperature $\mu$-metal shield surrounding the liquid helium dewar.  The SQUID and SQUID controller, with 100 kHz modulation frequency, were obtained from the company \emph{ezSQUID}.~\cite{EZSQUID}  A lock-in amplifier provides an ac source, and reads the feedback output from the SQUID controller as the resulting signal.

The samples were mounted on a sapphire rod using General Electric varnish, the long edge aligned with the direction of the applied magnetic field with an estimated accuracy of $\pm 2$ degrees.  The sapphire rod is clamped into the copper base of the sample holder and this sample holder is in turn heat sunk to the mixing chamber of a dilution refrigerator.

Frequency scans were taken by controlling the cryostat at a given temperature in the range of 500 to 1300 mK, and sweeping the frequency in the range of 1 mHz to 500 Hz.  It was ensured that, for all measurements presented here, the apparatus and samples were in thermal equilibrium.  Reproducibility, and the absence of further thermal relaxation, was verified by taking multiple scans separated by several hours.  The lower temperature limit of our data corresponds roughly to the point at which the maximum in $\chi''$ reaches our lowest measurement frequency of 1 mHz.  This temperature limit was chosen since much less information will be gained below that point without employing prohibitively low frequencies of measurement.  No problems of thermal equilibration were encountered that would have otherwise increased the base temperature of our measurement.  The excitation power was also varied by more than an order of magnitude to rule out heating of the sample or apparatus or other nonlinear effects.  Temperature scans of $\chi$ at fixed frequency were also obtained for four different measurement frequencies: 0.1, 1.2, 10 and 40 Hz.

\begin{figure}
\begin{center}
\includegraphics[width=3.25in,keepaspectratio=true]{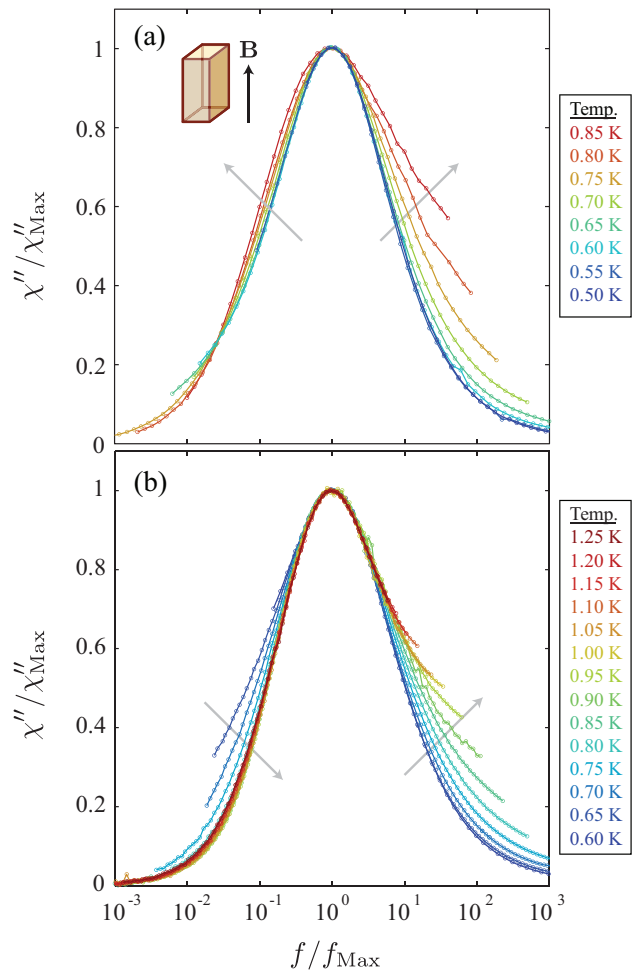}
\caption{(color online) Superimposed absorption spectra obtained by plotting $\chi''/\chi''_\mathrm{Max}$ against
$f/f_\mathrm{Max}$. (a) The absorption spectra taken on the less elongated sample (sample 2) of dimensions $1.1\times 1.1\times 2.6$ mm$^3$, \emph{before} correcting for the demagnetization effect.  The raw data, plotted in this way, exhibits a narrowing of the absorption spectrum with decreasing temperature. (b) Data fully corrected for the demagnetization effect, taken on a needle-shaped sample (sample 1).  The spectra change qualitatively, with a clear broadening of the low-frequency tail as the temperature is reduced.  However, the high-frequency tail continues to narrow with reducing temperature.  Above 1 K, the low-frequency tail also shows slight narrowing with reducing temperature, although the effect is rather subtle.  Arrows indicate increasing temperature.}
\label{chiNormComparisonFigure}
\end{center}
\end{figure}

\section{Results}

Results of ac susceptibility frequency scans, shown in Fig.~\ref{ChiNeedleFigure}, are qualitatively consistent with previous results taken on the spin ice material Dy$_2$Ti$_2$O$_7$.~\cite{Matsuhira2001,Snyder2001, Snyder2003, Snyder2004}  Slow relaxation is observed, characterized by broad absorption spectra $\chi''(f)$ and suppression or blocking of the in phase susceptibility $\chi'$ at higher frequencies.  Rapid freezing of the magnetic moments can be seen through the sharply dropping peak frequency, $f_\mathrm{Max}$, of the absorption spectrum $\chi''(f)$ as the temperature is lowered.   We see that by 500 mK, $f_\mathrm{Max}$ is already below our frequency window at less than 1 mHz.  At higher temperatures approaching 1.3 K, the peak position is beginning to plateau at around 100 Hz.  The magnitude of the susceptibility monotonically increases with decreasing temperature.  However, judging by the peak height of $\chi''$, the susceptibility appears to be leveling off at the lowest temperatures studied.

In order to make comparison of the spectra more amenable, we superimpose the spectra in Fig.~\ref{chiNormComparisonFigure}.  This is done by plotting data normalized by the peak susceptibility and frequency, so plotting $\chi''/\chi''_\mathrm{Max}$ against $f/f_\mathrm{Max}$.  Interestingly, if this is done before correcting for demagnetization with the less elongated sample, one finds a subtle narrowing of the absorption spectrum with lower temperatures, as shown in Fig.~\ref{chiNormComparisonFigure}(a).  This effect was previously noticed by Snyder \emph{et al.}\cite{Snyder2004} in measuring the susceptibility of Dy$_2$Ti$_2$O$_7$.

\begin{figure}
\begin{center}
\includegraphics[width=3.25in,keepaspectratio=true]{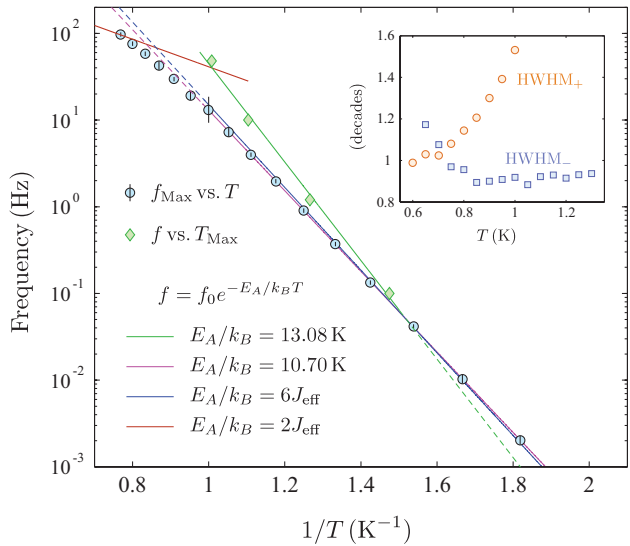}
\caption{ (color online) Frequency versus inverse temperature taken from this work and fits with different methods.  Green diamonds are obtained from the maxima $T_\mathrm{Max}(f)$ of temperature scans of $\chi''$.  The corresponding best fit Arrhenius law is shown as the green line.  The blue circles are obtained from the maxima of the absorption spectra, $f_\mathrm{Max}(T)$.  The best fit Arrhenius law, shown in magenta, gives $E_A/k_B = 10.70$ K, which is very close to $6J_\mathrm{eff}$ for Ho$_2$Ti$_2$O$_7$.  In fact, an Arrhenius law with $E_A/k_B = 6J_\mathrm{eff}$ also fits the data to within the uncertainty, as shown in blue.  The fits are plotted as solid lines in the range of the fitted data and extended with dashed lines.   The $2J_\mathrm{eff}$ Arrhenius law that should be expected in the plateau region is shown in red.}
\label{OurArrheniusPlots}
\end{center}
\end{figure}

Because of a large magnetic moment on the Ho$^{3+}$ sites, the susceptibility of Ho$_2$Ti$_2$O$_7$ is quite large and the demagnetization effect is crucially important to consider.  In fact, without correcting for demagnetization, one can obtain qualitatively different spectra due to mixing of $\chi'$ and $\chi''$.  If a demagnetization correction is performed, the narrowing behavior becomes less obvious, as seen in Fig.~\ref{chiNormComparisonFigure}(b).  The low-frequency tails of $\chi''(f)$ are found to broaden with decreasing temperature, whereas the high-frequency tails still narrow significantly with decreasing $T$.  Even in the demagnetization corrected data, there is some hint of the low-frequency side of the spectra beginning to broaden very slightly above 1 K.  Thus, the relaxation in spin ice is found to be noticeably different from the behavior of a standard glass or spin glass, where the absorption spectra show clear and largely symmetric broadening with reduced temperature.\cite{Reich1990, Quilliam2008}

Quantitatively, the broadening of the absorption spectra on a log scale can be parametrized by the half width at half maximum, either on the high frequency side (HWHM$_+$) or low frequency side (HWHM$_-$), both of which are plotted against temperature in the inset of Figure~\ref{OurArrheniusPlots}.  HWHM$_-$ is roughly constant at around 0.9 decades from 1.3 K down to 0.85 K (though showing barely discernible narrowing with lower $T$).  Below that point, it shows broadening, to almost 1.2 decades by 650~mK.  HWHM$_+$ shows a stronger and opposite temperature dependence, increasing from $\sim 1$ decade at the lowest temperatures studied here to 1.5 decades at 1.0~K.  These numbers can be compared to the 0.7 decades HWHM and symmetric absorption spectrum that results from a single energy barrier to relaxation.  The asymmetry of the spectra indicate that the relaxation is not described by a single characteristic Debye form.  Simple empirical forms such as the Cole-Cole,\cite{Cole1941} Davidson-Cole\cite{Davidson1950} and Havriliak-Negami\cite{Havriliak1967} functions are also not good fits to the data at low $T$, unlike what was found above 1.8 K in a previous analysis of the susceptibility of Dy$_2$Ti$_2$O$_7$.\cite{Matsuhira2001}

At low temperatures, the peak absorption frequency appears to approach an Arrhenius law, $f_\mathrm{Max} = f_0\exp(-E_A/T)$, thus resulting in a straight line when $\log f_\mathrm{Max}$ is plotted against $1/T$, as seen in Fig.~\ref{OurArrheniusPlots}.  Fitting the data below 1 K gives a best fit Arrhenius law with activation energy $E_A = 10.70$ K and $f_0 = 5.74\times 10^5$ Hz.  The resulting reduced chi-square statistic, $X^2$, is 0.12, implying that it is a very good fit to the data.  The value of $E_A$ so obtained is tantalizingly close to $6J_\mathrm{eff} = 10.80$~K for Ho$_2$Ti$_2$O$_7$.\cite{Melko2004}  In fact, an Arrhenius law with exactly $E_A = 6J_\mathrm{eff}$ is also a more than adequate fit with $f_0 = 6.57\times 10^5$ Hz and reduced $X^2 = 0.18$.  The low values of $X^2$ suggest that the fit is somewhat underconstrained.  Taking a 95\% confidence interval allows for $E_A = 10.70\pm 0.15$ K.   Above 1 K, the temperature dependence of the frequency $f_\mathrm{Max}$ becomes shallower than the Arrhenius-law fit, likely moving towards the quasi-plateau regime that was seen in DTO.\cite{Snyder2001,Snyder2004}  At the lowest measured temperatures, an onset of Arrhenius behavior was seen in DTO~\cite{Snyder2004} and in a later work a $\sim 6J_\mathrm{eff}$ Arrhenius law was shown to fit the lower end of that same experimental data.~\cite{Jaubert2010}  However, insufficient low temperature data has been obtained on DTO to be reasonably certain that the Arrhenius behavior develops as is seen here in HTO.

\begin{figure}
\begin{center}
\includegraphics[width=3.25in,keepaspectratio=true]{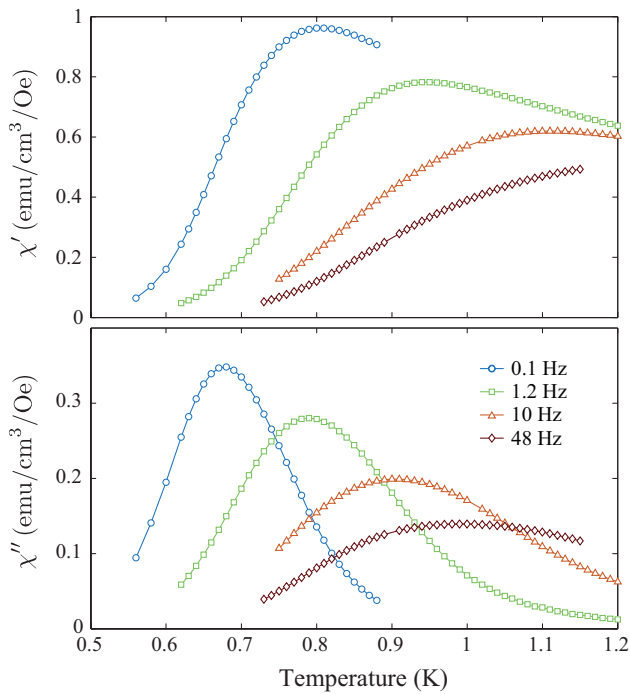}
\caption{(color online) Temperature scans of $\chi'$ and $\chi''$ of Ho$_2$Ti$_2$O$_7$ with the probe field aligned along [110], for four different measurement frequencies: 0.1 Hz, 1.2 Hz, 10 Hz and 48 Hz.  This data has been taken on a needle-shaped sample (sample 1) and has been corrected for the demagnetization effect.  Again, slow relaxation is observed, with a maxima in $\chi'$ or $\chi''$ moving to lower temperatures with lower frequency of measurement.}
\label{tempScansFigure}
\end{center}
\end{figure}

Detailed temperature scans at four different frequencies of measurement, 0.1 Hz, 1.2 Hz, 10 Hz, and 48 Hz, are shown in Fig.~\ref{tempScansFigure}.  Once again, the hallmarks of slow relaxation are observed, with suppression of $\chi'$ at low temperatures and a peak in $\chi''(T)$ near the inflection point of $\chi'(T)$.  The peak in $\chi''$ shifts to higher temperatures and broadens as the frequency of measurement, $f$, is increased.  The peak position of the temperature scans, or $T_\mathrm{Max}(f)$, may also be extracted in order to characterize the relevant time scale as a function of temperature.

\begin{figure}
\begin{center}
\includegraphics[width=3.25in,keepaspectratio=true]{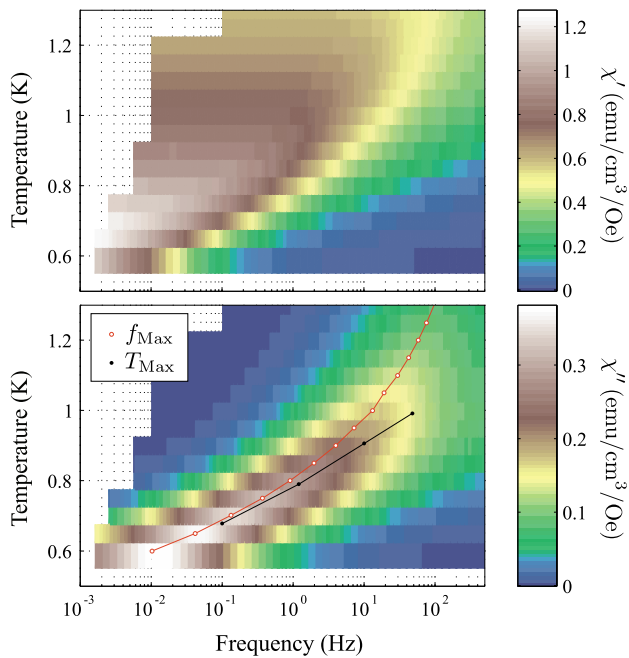}
\caption{(color online) Surface plot of $\chi'$ (top) and $\chi''$ (bottom) with temperature $T$ on the $y$-axis and measurement frequency $f$ on the $x$-axis.  Also shown, are the positions of the maxima in $\chi''$ either taken from temperature scans (black, filled circles) or frequency scans (red, open circles).}
\label{surfacePlotsFigure}
\end{center}
\end{figure}

It is found that the peak positions of the frequency scans and of the temperature scans lead to different functional forms, so that $f_\mathrm{Max}(T) \neq f(T_\mathrm{Max})$.  This is most easily demonstrated with surface plots of $\chi'(f,T)$ and $\chi''(f,T)$, shown in Fig.~\ref{surfacePlotsFigure}.  The maxima in $\chi''$ taken along the frequency axis and along the temperature axis are shown in unfilled and filled circles respectively and do not sit on the same curve.  While there is a very significant difference between these parametrizations at higher temperature, the two curves appear to be tending toward a single function at lower temperatures.  In previous work,\cite{Jaubert2009} $f_\mathrm{Max}(T)$ was used to compare with theory.  Ideally one would like to have theoretical calculations that can reproduce the entire $\chi(f,T)$ surfaces that are shown in Figure~\ref{surfacePlotsFigure}, but for now, $f_\mathrm{Max}(T)$ will serve as a good representative parameterization of our data.

The demagnetization \emph{correction} shifts the peak frequency significantly lower, especially at lower temperatures where the static susceptibility becomes quite large.  In other words, the demagnetization \emph{effect}, increases the measured relaxation times relative to the relaxation times that would be measured from the true, bulk susceptibility.   This is a result of $\chi'$ feeding into the measurement of $\chi''$.  This reduces the apparent absorption spectrum $\chi''_A$ at low frequencies where $\chi'$ is large, thereby pushing the peak frequency $f_\mathrm{Max}$ to higher frequency.  For more details on the demagnetization effect and correction, see the Appendix.


\section{Comparison with previous work}


Multiple experimental techniques have been used to study the relaxation of the three well established spin ice materials: Dy$_2$Ti$_2$O$_7$ (DTO),\cite{Snyder2004,Matsuhira2001,Lago2007} Ho$_2$Ti$_2$O$_7$ (HTO)\cite{Matsuhira2000,Clancy2009,Ehlers2003,Ehlers2004} and Ho$_2$Sn$_2$O$_7$ (HSO).\cite{Matsuhira2000}  In ac susceptibility, neutron spin echo, neutron scattering, and $\mu$SR relaxation, the following general features are observed.

Above $\sim 15$ K, one observes a high temperature regime where the transition rate of the individual spins is changing rapidly in $T$.  The energy scale for this regime appears to be set by the energy gap to the next excited crystal field energy level at $\sim 210$ K for DTO\cite{Snyder2001,Snyder2003,Snyder2004} and $\sim 290$ K for HTO.\cite{Ehlers2003} The Ho$^{3+}$ (or Dy$^{3+}$) moments have a truly Ising doublet ground state with matrix elements of the local $J_x$ and $J_y$ angular moment operators identically zero.  Transitions within this doublet are thus forbidden transitions.  Spin flips must therefore occur through either quantum tunneling (a slow process) or by passing through the excited crystal field energy levels (a much faster process).  Relaxation in the high temperature regime can thus be described by an Arrhenius law with an activation energy of $\sim 300$ K.  As the temperature is lowered it becomes difficult to populate the excited crystal field states reducing the transition rate.

Below $\sim 15$ K, the spins are very Ising-like and the materials enter a quasi-plateau regime where the tunneling rate levels off.  The vast majority of spin flips are now occurring via tunneling processes.  Note however, the tunneling is also accompanied via a change in energy resulting from monopole excitations, thus there remains a temperature dependence to the relaxation.  This plateau should correspond to the high temperature tail of a $2J_\mathrm{eff}$ Arrhenius behavior that results from the thermal excitation of independent (deconfined) monopole defects.\cite{Jaubert2009,Jaubert2010}  The monopole density in the plateau regime is high enough that there is significant screening of the magnetic charges to make the nearest neighbor spin ice model largely applicable.

Approaching $J_\mathrm{eff}$ (around 1.5 to 2 K),\cite{Bramwell2001} the system enters a regime exhibiting rapid freezing out of dynamics.\cite{Jaubert2010}  Here the time scales begin to increase dramatically as the system condenses into a disordered 2-in-2-out configuration.  Equivalently, it is now becoming thermally difficult to form monopole defects.  The monopole density decreases  which in turn reduces screening of the magnetic charges, making full treatment of the dipolar spin ice model necessary to describe the dynamics.\cite{Jaubert2009,Jaubert2010}

While different techniques and the different spin ice materials give quantitatively different time scales, they all agree on the qualitative behavior and on three clear regimes of relaxation.  For example, HTO neutron scattering experiments probing particular wave vectors resolve fluctuation rates $\sim 10$ ns, that are several orders of magnitude faster than those seen in the bulk susceptibility, with time scales around $\sim 1$ ms.\cite{Clancy2009}  Despite this significant difference in relaxation times, the same qualitative shape of $f(T)$ described above is observed.\cite{Clancy2009}  Neutron scattering measurements by Zhou \emph{et al.} also found a quantum tunnelling regime on the time scale of $\sim 0.01$~ns in the dynamic spin ice material Pr$_2$Sn$_2$O$_7$.\cite{ZhouPr2Sn2O72008}  Ehlers \emph{et al.} have found, using neutron spin echo experiments, the same three regimes of relaxation.\cite{Ehlers2003,Ehlers2004} Susceptibility measurements of DTO~\cite{Snyder2001} have found the time scale of the plateau to be on the order of $\sim 1$ ms, while $\mu$SR measurements of DTO have found the above qualitative shape, but with the time scale of the plateau relaxation on the order of 0.1 $\mu$s.~\cite{Lago2007}  This mismatch of timescales in the dynamics as determined by different measurement techniques remains an open question in the study of spin ice.\cite{Lago2007,Gardner2010}  The answer may lie in the wavevectors that are accessed by these various measurements; while neutron scattering accesses specific wavevectors, bulk susceptibility addresses $k=0$ and $\mu$SR is a local measurement, thus has contributions from all wavevectors.\cite{Gardner2010}

Interestingly, measurements of DTO by Orend\'{a}\v{c} \emph{et al.} using the magnetocaloric effect have accessed a frequency and temperature range lower than that of any ac susceptibility measurements.~\cite{Orendac2007}  Those measurements also seem to access a different time scale, showing both a higher rate of relaxation and a shallower temperature dependence of the freezing than what is observed in ac susceptibility.  Henceforth, we concentrate on comparing our data with that of other bulk susceptibility measurements of spin ice.


Even between ac susceptibility measurements, a direct comparison of data can be difficult to achieve.  For many of the measurements published on spin ice, a correction of the demagnetization effect has not been performed, or is not incorporated into the presented data.  Many of the previous results were obtained on powder samples, with which it is difficult to compare aligned single crystal measurements.  Of course different spin ice materials have different parameters in their respective Hamiltonians, which makes their quantitative behavior different.

\begin{figure}
\begin{center}
\includegraphics[width=3.25in,keepaspectratio=true]{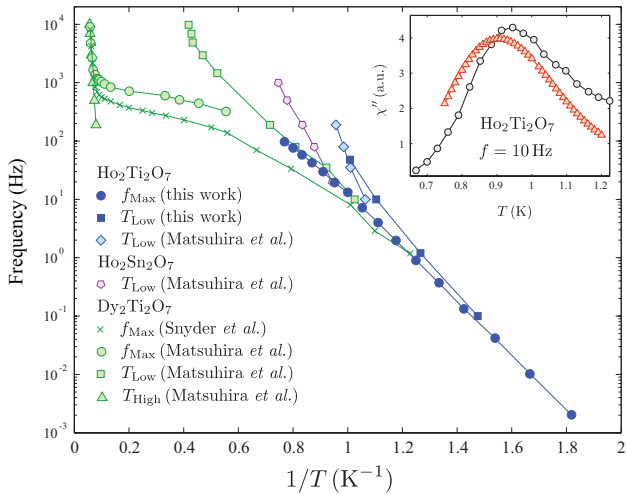}
\caption{(color online) Comparison of various ac susceptibility data on spin ice compounds.  Three methods for determining $f(T)$ are used.  Two different maxima may be determined from temperature scans of $\chi''$, which are labelled as $T_\mathrm{Low}$ and $T_\mathrm{High}$.  Only $T_\mathrm{Low}$ has been determined in our data because of the higher frequencies and temperatures required to measure $T_\mathrm{High}$.  Data on Dy$_2$Ti$_2$O$_7$ from Snyder \emph{et al.}\cite{Snyder2001} and from Matsuhira \emph{et al.}\cite{Matsuhira2001} are shown, and discrepancies can likely be attributed to demagnetization effects.  Data from Matsuhira \emph{et al.}\cite{Matsuhira2000} are shown on the materials Ho$_2$Ti$_2$O$_7$ and Ho$_2$Sn$_2$O$_7$ as well as our results on Ho$_2$Ti$_2$O$_7$.  The inset shows temperature scans (scaled arbitrarily for ease of comparison) from this work (red triangles) and from Ref.~\onlinecite{Matsuhira2000}, at the same frequency of 10 Hz.  The discrepancy could be attributed to a difference between powder samples\cite{Matsuhira2000} and the aligned single crystals measured here.}
\label{compilation}
\end{center}
\end{figure}

Finally, comparison is further hindered by the complex dynamical behavior of spin ice and the various measurement schemes and parametrizations that can be used.  One of the most obvious ways to parametrize the data is with the maximum of the absorption spectra at fixed temperature, or $f_\mathrm{Max}(T)$.  This parameter has been employed in several experimental works and in theory.\cite{Matsuhira2001,Snyder2004,Jaubert2009,Jaubert2010}  Alternatively, temperature scans at constant frequency may be employed and two maxima in $\chi''$ can be resolved at $T_\mathrm{Low}(f)$ and $T_\mathrm{High}(f)$.\cite{Matsuhira2000,Matsuhira2001}  The higher temperature maximum, $T_\mathrm{High}$, appears to be present only for frequencies above a certain threshold (near 200 Hz for DTO\cite{Matsuhira2001}).  While $T_\mathrm{High}$ has not been observed in HTO in zero field due to an unsuitable range of temperature and frequency, it has been observed, by Ehlers \emph{et al.}\cite{Ehlers2003, Ehlers2004} under an applied field of 1 T.  In temperatures scans, one may also extract maxima in $\chi'$.\cite{Shi2007}  In Figure~\ref{compilation}, we show a compilation of several of the previous results on DTO,\cite{Matsuhira2001,Snyder2001,Snyder2004} HTO\cite{Matsuhira2000} and HSO\cite{Matsuhira2000} compared with our results on HTO.

While qualitatively similar behavior is observed in ac susceptibility for all the materials studied, quantitative differences remain.  There is very little difference in the data below 1 K between our results on HTO and results on DTO from Refs.~\onlinecite{Snyder2001,Matsuhira2001}, despite the differences in Hamiltonian, specifically the effective n. n. exchange energy scale $J_\mathrm{eff}$, and in demagnetization effects.  We can draw from the physics of monopole excitations discussed in Refs.~\onlinecite{Castelnovo2008,Jaubert2009,Jaubert2010} to suggest what relations between the different relaxation rates should be expected.

First, the tunneling rates will be different in the two materials studied.  Tunneling rates for Ising moments like those in spin ice are functions of several different parameters.   The symmetry and strength of the crystal field, hence the energy of excited states above the Ising doublet, play an important role, giving the energy barrier through which the moments must tunnel.\cite{Snyder2001,Snyder2004}  Components of the exchange and dipolar interactions transverse to the local Ising direction result in mixing with the excited crystal field energy levels, permitting tunneling to occur.\cite{Schechter2008b,Quilliam2008}  Furthermore, nuclear hyperfine interaction strengths, very different between Dy$^{3+}$ and Ho$^{3+}$, can play a role in assisting or blocking spin flips.\cite{Schechter2008b,Quilliam2008}  To further complicate the system, we may anticipate that there will not be a single microscopic tunneling time, but a distribution.  However, spin ice has so far been treated with a single spin flip rate $f_0$.\cite{Jaubert2009,Jaubert2010}   In the regime where tunneling is dominant (from temperatures in the quasi-plateau regime and below), increasing the spin flip rate would be equivalent to stretching the frequency axis.  It is not clear at this time what the difference in tunneling rates is between the various spin ice materials, though our results suggest that they are at least in the same order of magnitude.

Within the nearest-neighbor spin ice (NNSI) model, the only other parameter governing the dynamics of the system would be $J_\mathrm{eff}$, or the energy barrier that must be overcome in order to excite monopole defects.\cite{Jaubert2009}  Changing $J_\mathrm{eff}$ would stretch the temperature axis, either elongating or shortening the plateau regime.   In Dy$_2$Ti$_2$O$_7$, $D_{nn} = 2.35$ K and $J_{nn} = -1.24$ K giving $J_\mathrm{eff} = 1.11$~K.\cite{Bramwell2001}  In Ho$_2$Ti$_2$O$_7$, the dipolar interaction is essentially the same, whereas $J_{nn} = -0.52$ K, giving a much higher value of $J_\mathrm{eff} = 1.83$ K.\cite{Melko2004}   Because of the higher $J_\mathrm{eff}$ in HTO, it enters the freezing, 2-in-2-out regime, at higher temperatures than DTO.\cite{Melko2004}

When one goes beyond the NNSI model into the dipolar spin ice model, the monopole defects begin to interact with a Coulomb law, $V(r_{ij}) = (\mu_0/4\pi) Q_iQ_j/r_{ij}$, as long as they occupy distant sites on the diamond lattice.  Thus a third parameter, the magnetic charge $Q$, starts to play an important role at lower temperatures as the monopole density is reduced and there is less screening of the magnetic charges.  The monopole charge is not that of the conjectured monopoles of particle physics, but is much smaller and has been determined theoretically\cite{Castelnovo2008} and experimentally\cite{Bramwell2009} to be
\begin{equation} Q = \pm\frac{2\mu}{a_d} =   \pm \left( \frac{32\pi D_{nn} a}{3\mu_0}\right)^{1/2} \end{equation}
where $\mu = 2g_J\mu_B\langle J_{\hat{z}} \rangle$ is the magnetic moment of the rare-earth ions, $a_d$ is the distance between adjacent sites of the diamond lattice and $a$ is the distance between spins on the pyrochlore lattice.  $Q$ is not directly related to $J_\mathrm{eff}$, but only the strength of the dipolar interaction $D_{nn}$.  If the NNSI model were to hold, a plot of $f_\mathrm{Max} \tau_0$ versus $T/J_\mathrm{eff}$ would in principle give the same result for all spin ice samples, from the quasi-plateau regime and below. Introduction of this third parameter, the monopole charge $Q$, should make such a simple scaling of relaxation curves impossible.

Surprisingly, despite all the differences between samples and measurements, our frequency scan maxima $f_\mathrm{Max}$ are quite close to those of Snyder \emph{et al.}\cite{Snyder2001,Snyder2004} on a poycrystalline sample of DTO.  While overlapping temperatures are not available, our results do seem to be approaching those of Matsuhira \emph{et al.}\cite{Matsuhira2001}, also on polycrystalline DTO, at higher temperatures.

We may more directly compare our temperature scans with the results of Ref.~\onlinecite{Matsuhira2000} where HTO has also been studied and a demagnetization correction has been applied.  Shown in the inset of Figure~\ref{compilation}, are our results at $f = 10$ Hz on a single crystal oriented along $[110]$ and those of Matsuhira \emph{et al.}\cite{Matsuhira2000} at the same frequency on a powder sample of HTO.  There is a clear difference between the curves, suggesting that the magnetic field orientation is an important parameter.  The relaxation times determined from $T_\mathrm{Low}(f)$ are found to have a steeper temperature dependence for the polycrystalline sample than for the $[110]$-oriented sample.  This would seem to suggest that energy barriers to the movement of monopole defects may be lower along the $[110]$ and symmetry related directions (in other words, along the so-called~\cite{Clancy2009} $\alpha$ or $\beta$ chains) than in other crystal directions.  In a perfect gas of interacting charges all directions are equivalent, but the magnetic monopoles in spin ice are confined to travel only in certain paths\cite{Jaubert2009,Jaubert2010} which could give rise to anisotropic behavior.

 Shi \emph{et al.},~\cite{Shi2007} have studied the relaxation of two DTO single crystals, one oriented along the $[111]$ direction and the other oriented along the $[100]$ direction.  They note that there is a difference between $T_f(f)$ (the maximum in $\chi'(T)$) between these two scenarios.  However, it is not clear that their data has been corrected for demagnetization.  The two samples have different geometries and the different orientations will give rise to different magnitudes of susceptibility since the magnetic field will be differently aligned with the different basis spins.  This change in magnitude will heavily influence the apparent susceptibility $\chi_A(f,T)$.

\section{Conclusions}

To conclude, we have performed a careful study of the frequency dependent spin freezing in a single crystal of the dipolar spin ice material \HTO at low temperatures.  The results are qualitatively consistent with research by other groups on Ho$_2$Ti$_2$O$_7$,\cite{Matsuhira2000} on the related materials Dy$_2$Ti$_2$O$_7$\cite{Matsuhira2001,Snyder2001,Snyder2004} and Ho$_2$Sn$_2$O$_7$\cite{Matsuhira2001} and also with theoretical work that relates such dynamics to the motion of magnetic monopole excitations.\cite{Jaubert2009,Jaubert2010}  The results presented here represent an exploration much deeper into the Coulomb phase of spin ice, with much lower temperatures and frequencies of measurement than were employed for ac susceptibility measurements previously.

One of the most striking aspects of our results is the observation of a seemingly temperature activated regime below 1~K where the dynamics are well fit by an Arrhenius law, consistent with a $6J_\mathrm{eff}$ activation energy.  It has also been noted\cite{Jaubert2010} that an Arrhenius law with an activation energy close to $6J_\mathrm{eff}$ is consistent with the low temperature limit of measurements on Dy$_2$Ti$_2$O$_7$.\cite{Snyder2001,Snyder2004}  While very few points contribute to that conclusion for DTO, it nonetheless supports our observation of such an effect in HTO.  A more thorough investigation of the low temperature behavior of DTO with full demagnetization correction would be valuable to verify or disprove the existence of the simple law that we have observed.

It is not clear what could give rise to such a simple temperature activated regime in dipolar spin ice, which seems at odds with the current theoretical treatment of magnetic monopoles in dipolar spin ice.\cite{Jaubert2009,Jaubert2010} Such theory instead predicts ever increasing activation energies as monopole density and thus screening is reduced.\cite{Jaubert2009,Jaubert2010}  Furthermore, a one-parameter ($J_\mathrm{eff}$) fit to this low temperature regime surprisingly suggests that the monopole charge $Q$ might be an irrelevant parameter. There remains the possibility that the accessible temperature range is simply narrow enough that a more complex $f(1/T)$ function only appears to follow a linear behavior.  The Arrhenius law holds well over almost 4 orders of magnitude in frequency, though applies only over a factor of 2 in temperature.

Our measurements have also illustrated the complex nature of the spectra.  While qualitatively glassy relaxation is seen, the precise behavior of the susceptibility spectra is quite different from that of spin glasses.\cite{Reich1990,Quilliam2008}  In particular, as the temperature is lowered, the high frequency tail is found to narrow appreciably where the low frequency tail broadens.  Clearly illustrated here is the extreme importance of the demagnetization correction for such materials and its impact on the shape of the absorption spectra, for example.

When compared to experiments on Dy$_2$Ti$_2$O$_7$, theory currently shows a persistent mismatch at lower temperatures, with the experiments showing longer relaxation times than would be expected theoretically.\cite{Jaubert2010}  While those experiments have not taken into account the demagnetization effect, this will not help explain the discrepancy and should in fact make it worse as correcting for demagnetization will lower the peak frequencies.

In order to resolve such discrepancies, it has been suggested\cite{Jaubert2010} that higher order terms, such as next nearest neighbor exchange, $J_2$, may need to be added to the currently accepted Hamiltonian of the dipolar spin ices.  Such interactions were already found to be quite important for matching theory to experiment in DTO.\cite{Yavorskii2008}  It would also be useful to consider the possibility that the low temperature susceptibility of \HTO is affected by bound pairs of monopoles.  Such very slowly relaxing pairs of monopoles have recently been predicted to occur after thermally quenching spin ice.\cite{Castelnovo2010}  It is not clear, however, whether experimentally inaccessible cooling rates are required to form these pairs or whether they are practically unavoidable in the real systems.

It is hoped that the very exciting theory of monopole excitations in spin ice can begin to explain some of the peculiar effects that are seen in these susceptibility measurements.  Equivalently, these results, having been performed with careful demagnetization correction on single crystal samples and to comparatively low frequencies and temperatures, should serve as an important benchmark for testing the theory of monopole physics in spin ice.

\begin{acknowledgments}
We acknowledge A. B. Dabkowski, E. Mazurek, J.~P.~C. Ruff and S. Selesnic for contributions to the crystal growth and preparation, as well as very informative conversations with L.~Jaubert.  We thank M.~J.~P.~Gingras for many valuable discussions and for a critical reading of the manuscript.  We also acknowledge the contributions of S. Meng and C. G. A. Mugford to the measurement apparatus.  Funding for this research was provided by NSERC, CFI, OFI, MMO and the Research Corporation.
\end{acknowledgments}

\appendix*
\section{Calibration and Demagnetization Correction}

\begin{figure}
\begin{center}
\includegraphics[width=3.25in,keepaspectratio=true]{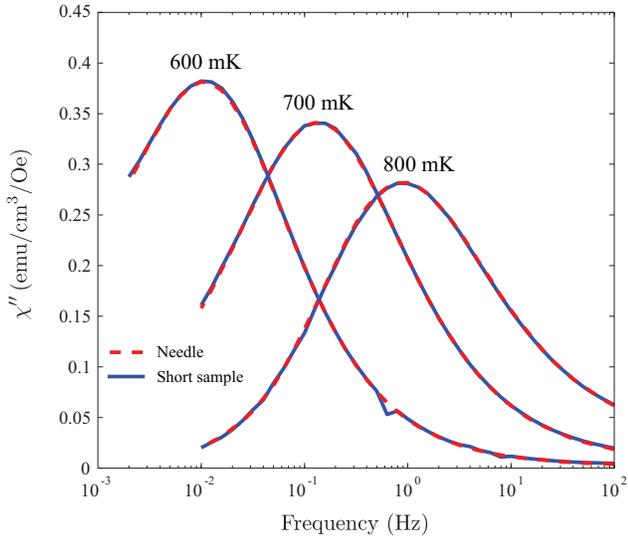}
\caption{(color online) Comparison of the calibrated and demagnetization corrected susceptibility of the two sample geometries studied in this work.  The needle-shaped sample with demagnetization factor $N_1$ is shown with a dashed red line and the less elongated sample with demagnetization factor $N_2$ is shown as a solid blue line.  Excellent agreement between the two data sets is obtained after suitable selection of the calibration factors.   While overlap is determined from the absorption spectra $\chi''$, the in-phase susceptibility $\chi'$ is also found to match very well.}
\label{calibrationFigure}
\end{center}
\end{figure}

Because of the large susceptibility of Ho$_2$Ti$_2$O$_7$, the demagnetization correction is quite significant and must be dealt with quite carefully in order to obtain accurate results.  Not accounting for the demagnetization effect can even result in \emph{qualitatively} different susceptibility spectra.  Here we use the standard relation
	\begin{equation}
	\frac{1}{\chi} = \frac{1}{\chi_A} - 4\pi N
	\label{DemagEffect}
	\end{equation}
to perform the demagnetization correction, where $\chi$ is the true complex susceptibility of the sample, $\chi_A$ is the apparent, or measured susceptibility, and $N$ is the demagnetization factor.  Because $\chi = \chi' - i\chi''$ is complex and has a great deal of frequency dependence, there is significant mixing between $\chi'$ and $\chi''$.  This alters the shape of the spectra obtained.  Thus it can be seen from expanding Eq.~\ref{DemagEffect}, that
	\begin{equation}
	\chi' = \frac{ \chi'_A - 4\pi N( \chi'^2_A + \chi''^2_A) }{ (1 - 4\pi N\chi'_A)^2 + (4\pi N\chi''_A)^2}
	\end{equation}
	\begin{equation}
	\chi'' = \frac{ \chi''_A}{  (1 - 4\pi N\chi'_A)^2 + (4\pi N\chi''_A)^2 }
	\end{equation}	
as discussed in Ref.~\onlinecite{Matsuhira2001}, for example.

\begin{figure}
\begin{center}
\includegraphics[width=3.25in,keepaspectratio=true]{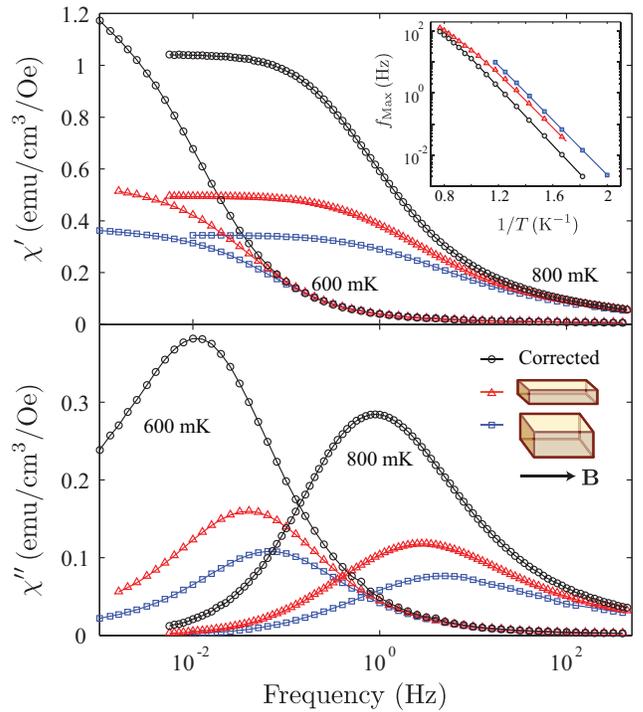}
\caption{(color online) The effects of the demagnetization effect on the susceptibility of Ho$_2$Ti$_2$O$_7$ at two different temperatures, 600 and 800 mK.  Blue and red curves are uncorrected data (apparent susceptibility) measured on a short sample of dimensions $1.1\times 1.1\times2.6$ mm$^3$ and on a needle shaped sample of dimensions $0.6\times 0.6\times 3$ mm$^3$, respectively.  The black curves are data that has been fully corrected for the demagnetization effect.  As can be seen from the figure, the correction is very significant, more than doubling the susceptibility at low temperatures and low frequencies where it is most important.  The inset shows the maximum frequencies $f_\mathrm{Max}$ obtained before (blue and red) and after (black) correcting for demagnetization, as a function of inverse temperature.}
\label{demagFigure}
\end{center}
\end{figure}

While one method of calibration could involve the use of a calibration standard, such as a perfectly diamagnetic superconducting sample, this method is fraught with difficulties.  To be completely accurate, one would need a calibration standard that is exactly the same size and shape and positioned in the same place as the sample.

Our method, instead, involves the measurement of two different sample geometries with demagnetization factors $N_1$ and $N_2$ respectively.  The raw susceptibility signal of each sample geometry is taken in arbitrary units, as $V_1$ and $V_2$ respectively, over a common range of temperatures and frequencies.  An array of trial calibration factors $C_1$ and $C_2$ for each of the samples is generated.  For each entry, the apparent susceptibilities, $\chi_{A1} = C_1V_1$ and $\chi_{A2} = C_2V_2$, are corrected for demagnetization according to Eq.~\ref{DemagEffect}.  An overlap is determined from the absorption spectrum data points with the parameter $\sum (\chi''_1 - \chi''_2)^2 / (\chi_1''^2 + \chi_2''^2)$ and the calibration factors $C_1$ and $C_2$ are chosen in order to minimize that parameter (maximize the overlap) or so $\chi_1 = \chi_2 = \chi$ for all data points.

This method utilizes a calibration of the sample which ensures proper scaling of the apparent susceptibility with sample geometry, thereby providing a very accurate correction of the demagnetization effect.  The only disadvantage, besides increased measurement time, is that it entails the assumption that the true susceptibility of the samples $\chi$ is not sample geometry dependent and that only the apparent susceptibility $\chi_A$ changes.  It is likely that, apart from extreme situations like a thin film of material, the shape of the sample does not affect intrinsic thermodynamic quantities such as the true magnetic susceptibility.

Specifically, we have used, for our calibration, frequency scans of the susceptibility at three different temperatures: 600, 700 and 800 mK.  The most needle-shaped sample geometry has dimensions $0.6\times0.6\times 3$ mm$^3$ and therefore has a demagnetization factor of $N_1 = 0.085$.  The other, less elongated but larger sample has dimensions $1.1\times 1.1\times 2.6$ mm$^3$ with demagnetization factor $N_2 = 0.171$.  The demagnetization correction was calculated with the analytical form for a rectangular prism given in Ref.~\onlinecite{Aharoni1998}.   The excellent overlap so-obtained is shown in Figure~\ref{calibrationFigure}.  The effects of sample geometry on the apparent susceptibility spectra and on the peak frequencies, $f_\mathrm{Max}$, compared with the fully corrected data, are shown in Figure~\ref{demagFigure}.

The importance of the demagnetization effect has not always been highlighted in other measurements of spin ice though in some cases, it has been pointed out to be quite important.\cite{Bramwell2000,Morris2009}  In measurements of polycrystalline HTO, by Matsuhira \emph{et al.}, a demagnetization correction was performed.\cite{Matsuhira2000}  While results of Matsuhira \emph{et al.} on DTO were corrected for demagnetization before the final analysis, much of the published data with which we might compare our results are left uncorrected.\cite{Matsuhira2001}  Recent magnetization measurements\cite{Morris2009} were also corrected for demagnetization, particularly important since they were performed on a complicated sample geometry in large magnetic field.  In some works, no mention was made as to whether a demagnetization correction was used.\cite{Snyder2001, Snyder2003, Snyder2004, Shi2007}    Here we have measured a single crystal with a well-defined and calculable\cite{Aharoni1998} demagnetization factor $N$ and have performed a careful correction of the demagnetization effect.  We can thus say with confidence that we are measuring the true, bulk susceptibility of Ho$_2$Ti$_2$O$_7$.

\bibliographystyle{h-physrev3}

\begin{thebibliography}{10}

\bibitem{Castelnovo2008}
C.~Castelnovo, R.~Moessner, and S.~L. Sondhi,
\newblock Nature {\bf 451}, 42 (2008).

\bibitem{Bramwell2009}
S.~T. Bramwell {\em et~al.},
\newblock Nature {\bf 461}, 956 (2009).

\bibitem{Morris2009}
D.~J.~P. Morris {\em et~al.},
\newblock Science {\bf 326}, 411 (2009).

\bibitem{Fennell2009}
T.~Fennell {\em et~al.},
\newblock Science {\bf 326}, 415 (2009).

\bibitem{Kadowaki2009}
H.~Kadowaki {\em et~al.},
\newblock J. Phys. Soc. Japan {\bf 78}, 103706 (2009).

\bibitem{Jaubert2009}
L.~Jaubert and P.~Holdsworth,
\newblock Nature Phys. {\bf 5}, 258 (2009).

\bibitem{Jaubert2010}
L.~D.~C. Jaubert and P.~C.~W. Holdsworth,
\newblock arXiv:1010.0970 .

\bibitem{Wang2006}
R.~F. Wang {\em et~al.},
\newblock Nature {\bf 439}, 303 (2006).

\bibitem{Ladak2010}
S.~Ladak, D.~E. Read, G.~K. Perkins, L.~F. Cohen, and W.~R. Branford,
\newblock Nature Physics {\bf 6}, 359 (2010).

\bibitem{Mengotti2010}
E.~Mengotti {\em et~al.},
\newblock Nature Phys. {\bf 7}, 68 (2010).

\bibitem{Morgan2010}
J.~P. Morgan, A.~Stein, S.~Langridge, and C.~H. Marrows,
\newblock Nature Phys. {\bf 7}, 75 (2010).

\bibitem{Snyder2001}
J.~Snyder, J.~S. Slusky, R.~J. Cava, and P.~Schiffer,
\newblock Nature {\bf 413}, 48 (2001).

\bibitem{Snyder2003}
J.~S. Snyder {\em et~al.},
\newblock Phys. Rev. Lett. {\bf 91}, 107201 (2003).

\bibitem{Snyder2004}
J.~Snyder {\em et~al.},
\newblock Phys. Rev. B {\bf 69}, 064414 (2004).

\bibitem{Matsuhira2001}
K.~Matsuhira, Y.~Hinatsu, and T.~Sakakibara,
\newblock J. Phys.: Condens. Matter {\bf 13}, L737 (2001).

\bibitem{Matsuhira2000}
K.~Matsuhira, Y.~Hinatsu, K.~Tenya, and T.~Sakakibara,
\newblock J. Phys.: Condens. Matter {\bf 12}, L649 (2000).

\bibitem{Shi2007}
J.~Shi {\em et~al.},
\newblock J. Magn. Magn. Mater. {\bf 310}, 1322 (2007).

\bibitem{Ehlers2004}
G.~Ehlers {\em et~al.},
\newblock J. Phys.: Condens. Matter {\bf 16}, S635 (2004).

\bibitem{Cornelius2001}
A.~Cornelius and J.~Gardner,
\newblock Phys. Rev. B {\bf 64}, 060406 (2001).

\bibitem{Harris1997}
M.~J. Harris, S.~T. Bramwell, D.~F. McMorrow, T.~Zeiske, and K.~W. Godfrey,
\newblock Phys. Rev. Lett. {\bf 79}, 2554 (1997).

\bibitem{Ramirez1999}
A.~Ramirez, A.~Hayashi, R.~Cava, R.~Siddharthan, and B.~Shastry,
\newblock Nature {\bf 399}, 333 (1999).

\bibitem{denHertog2000}
B.~C. den Hertog and M.~J.~P. Gingras,
\newblock Phys. Rev. Lett. {\bf 84}, 3430 (2000).

\bibitem{Higashinaka2003}
R.~Higashinaka, H.~Fukazawa, and Y.~Maeno,
\newblock Phys. Rev. B {\bf 68}, 014415 (2003).

\bibitem{Hiroi2003}
Z.~Hiroi, K.~Matsuhira, S.~Takagi, T.~Tayama, and T.~Sakakibara,
\newblock J. Phys. Soc. Japan {\bf 72}, 411 (2003).

\bibitem{Pauling}
L.~Pauling,
\newblock {\em The Nature of the Chemical Bond} (Cornell University Press,
  Ithica, New York, 1945).

\bibitem{Gingras2001}
M.~J.~P. Gingras and B.~C. den Hertog,
\newblock Can. J. Phys {\bf 79}, 1339 (2001).

\bibitem{Melko2001}
R.~G. Melko, B.~C. den Hertog, and M.~J. Gingras,
\newblock Phys. Rev. Lett. {\bf 87}, 067203 (2001).

\bibitem{Gardner2010}
J.~Gardner, M.~Gingras, and J.~E. Greedan,
\newblock Rev. Mod. Phys. {\bf 82}, 53 (2010).

\bibitem{Fukazawa2002}
H.~Fukazawa, R.~G. Melko, R.~Higashinaka, Y.~Maeno, and M.~J.~P. Gingras,
\newblock Phys. Rev. B {\bf 65}, 054410 (2002).

\bibitem{Melko2004}
R.~G. Melko and M.~J.~P. Gingras,
\newblock J. Phys.: Condens. Matter {\bf 16}, R1277 (2004).

\bibitem{Balents2010}
L.~Balents,
\newblock Nature {\bf 464}, 199 (2010).

\bibitem{Clancy2009}
J.~P. Clancy {\em et~al.},
\newblock Phys. Rev. B {\bf 79}, 014408 (2009).

\bibitem{Gardner1998}
J.~S. Gardner, B.~D. Gaulin, and D.~{McK. Paul},
\newblock J. Cryst. Growth {\bf 191}, 740 (1998).

\bibitem{EZSQUID}
SQUIDs and controller obtained from Michael M\"{u}ck, EZ-SQUID.

\bibitem{Reich1990}
D.~H. Reich {\em et~al.},
\newblock Phys. Rev. B {\bf 42}, 4631 (1990).

\bibitem{Quilliam2008}
J.~A. Quilliam, S.~Meng, C.~G.~A. Mugford, and J.~B. Kycia,
\newblock Phys. Rev. Lett. {\bf 101}, 187204 (2008).

\bibitem{Cole1941}
K.~S. Cole and R.~H. Cole,
\newblock J. Chem. Phys. {\bf 9}, 341 (1941).

\bibitem{Davidson1950}
D.~W. Davidson and R.~H. Cole,
\newblock J. Chem. Phys. {\bf 18}, 1417 (1950).

\bibitem{Havriliak1967}
S.~Havriliak and S.~Negami,
\newblock Polymer {\bf 8}, 161 (1967).

\bibitem{Lago2007}
J.~Lago, S.~Blundell, and C.~Baines,
\newblock J. Phys.: Condens. Matter {\bf 19}, 326210 (2007).

\bibitem{Ehlers2003}
G.~Ehlers {\em et~al.},
\newblock J. Phys.: Condens. Matter {\bf 15}, L9 (2003).

\bibitem{Bramwell2001}
S.~T. Bramwell {\em et~al.},
\newblock Phys. Rev. Lett. {\bf 87}, 047205 (2001).

\bibitem{ZhouPr2Sn2O72008}
H.~D. Zhou {\em et~al.},
\newblock Phys. Rev. Lett. {\bf 101}, 227204 (2008).

\bibitem{Orendac2007}
M.~Orend\'{a}\v{c} {\em et~al.},
\newblock Phys. Rev. B. {\bf 75}, 104425 (2007).

\bibitem{Schechter2008b}
M.~Schechter,
\newblock Phys. Rev. B {\bf 77}, 020401 (2008).

\bibitem{Yavorskii2008}
T.~{Yavors'kii}, T.~Fennell, M.~J.~P. Gingras, and S.~T. Bramwell,
\newblock Phys. Rev. Lett. {\bf 101}, 037204 (2008).

\bibitem{Castelnovo2010}
C.~Castelnovo, R.~Moessner, and S.~L. Sondhi,
\newblock Phys. Rev. Lett. {\bf 104}, 107201 (2010).

\bibitem{Aharoni1998}
A.~Aharoni,
\newblock J. Appl. Phys. {\bf 83}, 3432 (1998).

\bibitem{Bramwell2000}
S.~T. Bramwell, M.~N. Field, M.~J. Harris, and I.~P. Parkin,
\newblock J. Phys.: Condens. Matter {\bf 12}, 483 (2000).

\end{thebibliography}

\end{document}